\documentclass[prl,twocolumn,amsmath,aps,tightenlines,floatfix]{revtex4}
\usepackage{latexsym,amssymb,amsmath}
\usepackage{epsfig}
\usepackage{subfigure}
\usepackage{citesort}
\usepackage[lflt]{floatflt}

\usepackage{graphicx}
\usepackage{dcolumn}
\usepackage{subfigure}
\usepackage{bm}

\newcommand{\bra}[1]{ \langle #1 |}
\newcommand{\ket}[1]{ | #1 \rangle }

\newcommand{\expect}[1]{ \langle #1 \rangle}
\def\simlt{\mathrel{\lower .3ex \rlap{$\sim$}\raise .5ex \hbox{$<$}}}
\def\simgt{\mathrel{\lower .3ex \rlap{$\sim$}\raise .5ex \hbox{$>$}}}

\begin{document}
\title{Dynamics of a Complex Quantum Magnet}
\vspace{1cm}

\author{James W. Landry}
   \email{jwlandr@sandia.gov}
   \affiliation{Sandia National Laboratory,
   Albuquerque, NM  87185}
\author{S.N. Coppersmith}
   \email{snc@physics.wisc.edu}
   \affiliation{Physics Department, University of Wisconsin,
   1150 University Ave., Madison, WI  53706}

\date{\today}

\begin{abstract}
We have computed the low energy quantum states
and low frequency dynamical susceptibility of complex
quantum spin systems in the limit of strong interactions,
obtaining exact results for system sizes
enormously larger than accessible previously.
The ground state is a complex superposition of a substantial
fraction of all the classical ground states, and yet the dynamical
susceptibility exhibits sharp resonances reminiscent of the behavior
of single spins.  These results show that strongly interacting quantum
systems can organize to generate coherent excitations and shed
light on recent experiments demonstrating that coherent
excitations are present in a disordered spin liquid~\cite{ghosh2002}.
The dependence of the energy spectra on system size differs
qualitatively from that of the energy spectra of random
undirected bipartite graphs with similar statistics, implying that
strong interactions are giving rise to these unusual spectral
properties.

\end{abstract}

\pacs{75.50.Lk, 75.10.Nr}

\maketitle

The motivation for understanding the dynamics of quantum systems
continues to grow as technology miniaturizes.  That quantum and
classical dynamics differ fundamentally is underscored by the
indications that quantum dynamics can be harnessed in the form of
quantum computation to solve problems that are intractable using the
processes of classical physics~\cite{quantumcomputing}.

The promise and difficulty of quantum dynamics both arise because the
amount of information needed to specify a quantum state grows
exponentially with the system size.  The number of complex numbers
needed to specify a general quantum state of $N$ quantum-mechanical
two-state spins (or qubits) is $2^{N}$.  A classical computer performing
$10^{14}$ floating point operations per 
second~\cite{fastestcomputernote} would take $10^8$ years to perform a
single operation when $N=100$ and $10^{21}$ years when $N=144$.

Here we study a frustrated spin model that is a quantum generalization
of the two-dimensional $\pm J$ Edwards-Anderson (E-A) spin glass
model~\cite{EdwardsMay1975}, a canonical example of a classical system
whose competing interactions give rise to many low-energy states.  The
essential physical ingredients of the E-A model arise in a wide variety
of optimization problems in many fields~\cite{mezard87}: the system
cannot satisfy simultaneously all its constraints, and many different
configurations are equally effective in minimizing the energy.  Though
the two-dimensional $\pm J$ E-A model is simpler than the
three-dimensional version---it is disordered at all nonzero
temperatures~\cite{PalassiniApr2001} and individual ground states can be
found in a time that scales as a polynomial of the system
size~\cite{BiecheAug1980}---it has a large ground state degeneracy and a
complex energy
landscape~\cite{PalassiniApr2001,saul_and_kardar,santoro2002}.  Motivated
by recent experiments on ${\rm LiY_xHo_{1-x}F_4}$ that demonstrate that
quantum tunneling has profound effects on the dynamics of a
three-dimensional Ising spin glass with dipolar
couplings~\cite{brooke99,brooke2001,ghosh2002}, we study the quantum
system obtained by adding a small quantum tunneling term to the
two-dimensional E-A model.

We calculate numerically exact eigenstates and dynamical response
functions of quantum E-A models with up to $144$ spins in the limit of
strong interaction strength.  At low frequencies the excitation spectrum
is remarkably sparse.  The number of excitations per unit energy
at low energies does not vary significantly
with system size, even though the number of eigenvalues grows
exponentially with system size while the energy bandwidth is only
growing polynomially.  The eigenvalues and eigenvectors of each
realization are obtained by diagonalizing the adjacency
matrix of a graph, and we demonstrate that strong correlations play a
vital role by comparing the spin glass excitation spectra
to spectra of graphs with similar connectivity
statistics but randomly chosen connections.  These theoretical results
provide a natural framework for understanding recent experiments
demonstrating the presence of sharp, saturable resonances in the quantum
spin liquid ${\rm LiY_{0.955}Ho_{0.045}F_4}$~\cite{ghosh2002}.

We study two-dimensional systems with periodic boundary conditions
in which spin-$1/2$ spins interact
with nearest neighbors on
$\sqrt{N} \times \sqrt{N}$ square lattices.
The quantum Hamiltonian is
\begin{equation}
H_Q = -\sum_{\langle ij \rangle} J_{ij} \sigma_{i,z} \sigma_{j,z}
+ \Gamma \sum_i \sigma_{i,x}~,
\label{eq:quantum_model}
\end{equation}
where $\sigma_{i,x}$ and $\sigma_{i,z}$ are Pauli matrices:
$\sigma_{i,x} = \begin{pmatrix} 0 & 1 \\ 1 & 0 \end{pmatrix}$ and
$\sigma_{i,z} = \begin{pmatrix}1 & 0 \\ 0 & -1 \end{pmatrix}$.
The sum  $\langle ij \rangle$ is over all nearest neighbor pairs.
A given sample has a fixed realization of bonds in which
each bond $J_{ij}$ is chosen to be $-J$
and $+J$ with equal probability.
This Hamiltonian, which has been studied by many
groups~\cite{transverseIsingspinglassrefs},
is believed to be relevant to the experimental system
${\rm LiY_{x}Ho_{1-x}F_4}$~\cite{brooke99,brooke2001,ghosh2002},
where the coefficient $\Gamma$ is tunable because it is proportional
to an applied transverse magnetic field.
The $\Gamma \rightarrow 0$ limit is
the two-dimensional $\pm J$ Edwards-Anderson (E-A) 
model~\cite{EdwardsMay1975}.

As $\Gamma/J \rightarrow 0$
the low energy quantum states $\ket{\psi_n}$ become superpositions
of states corresponding to
ground state configurations $\ket{\alpha}$ of the classical model,
\begin{equation}
\ket{\psi_n} = \sum_\alpha c_{\alpha n}\ket{\alpha}~.
\label{eq:c_definition}
\end{equation}
The number of ground states of the classical
two-dimensional $\pm J$ E-A model grows with
$N$ much more slowly than the number of configurations
(though still exponentially)~\cite{saul_and_kardar,Landry2001a};
we use an algorithm that
we have developed that finds all the
classical ground states
efficiently~\cite{Landry2001a}
and standard degenerate perturbation theory~\cite{shankar80}
to compute the low energy quantum states of about
$90\%$ of $10 \times 10$ realizations
and about $40\%$ of
$12 \times 12$ realizations~\cite{limitation_note}.
In the limit $\Gamma/J \rightarrow 0$,
each low energy quantum eigenstate is a
superposition of classical ground state configurations related to each
other by serial flipping of individual flippable spins, where a
flippable spin is one with an equal number of satisfied and
unsatisfied bonds~\cite{CoppersmithOct1991}.
This follows because
the matrix element $\bra{\alpha}\left
(\sum_i\sigma_{ix}\right )\ket{\beta}$ of the quantum tunneling term
between any two classical configurations corresponding to the
quantum basis states $\ket{\alpha}$
and $\ket{\beta}$ is nonzero only if the two states differ by a
single spin flip, and, to lowest order, only states $\ket{\alpha}$ and
$\ket{\beta}$ with the same energy contribute.
The number of flippable spins clearly is no greater
than $N$, the number of spins in the system, so
the matrix 
characterizing the possible transitions between classical ground
states is extremely sparse, and thus well-suited for diagonalization
using Lanczos techniques~\cite{cullum85,dagotto94,lanczos96}.
We have computed the dynamical magnetic susceptibility by finding
low-energy eigenvalues and eigenstates
using the sparse matrix ARPACK numerical library~\cite{Arpack2001}
with C++ bindings~\cite{Shaw2001}.
Most of the results shown here are for individual realizations;
though there are large sample-to-sample variations in the
number of ground states for a given system size~\cite{Landry2001a},
the qualitative results on which we focus are robust.

The ground state dynamical magnetic susceptibility $\chi^{''}(\omega)$
characterizes the response of a system at zero temperature to a
magnetic field applied along the $z$ axis
oscillating at frequency $\omega$~\cite{susceptibility2}.
The susceptibility consists of sets of Dirac $\delta$-function
peaks (that in physical systems
spread out into a finite width in frequency due to decoherence
processes);
each peak occurs at a frequency that is $\hbar$ times the
energy difference between an excited state and the ground state.
For the $\pm J$ spin glass, as
$\Gamma/J \rightarrow 0$ the value of $J$ affects only the energy zero and
the susceptibility at frequencies $\omega$ satisfying $\hbar\omega \ll J$
depends only on the ratio $\hbar \omega/\Gamma$.

\begin{figure}
\begin{center} 
\includegraphics[clip,width=0.72\hsize]%
{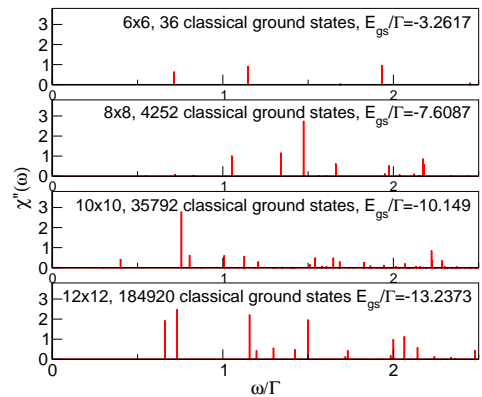}
\vskip .2cm
\caption{Zero temperature dynamic magnetic susceptibility
$\chi^{''}(\omega)$ of systems of size $6 \times 6$,
$8 \times 8$, $10 \times 10$ and $12 \times 12$.
The peaks in the susceptibility occur at frequencies $\omega$
that satisfy $\hbar \omega = E_n-E_{0}$, where $E_n$ is the
energy of an excited state and $E_{0}$ is the energy of
the ground state.  The density of low-energy excitations
does not increase appreciably as the system size increases,
even though an exponentially increasing number of states
are in an energy bandwidth that grows
approximately linearly with the system size.}
\label{fig:susceptibility}
\end{center}
\end{figure}

Figure~\ref{fig:susceptibility} shows the zero temperature
dynamic magnetic susceptibility of systems of size $6 \times 6$,
$8 \times 8$, $10 \times 10$, and $12 \times 12$.
The density of low energy excitations increases extremely slowly
with system size.  This result is surprising because the number of
energy eigenvalues grows exponentially with the number of spins
$N$, while these energies all lie within a bandwidth that grows
roughly linearly with $N$.

\begin{figure}
\begin{center}
\includegraphics[clip,width=0.72\hsize]%
{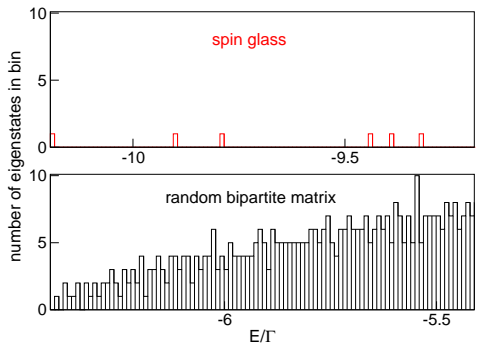}
\vskip 0.2cm
\caption{Density of states as a function of energy
at low energies for the largest connected component (with 
17040 nodes and 77684 links) of the
graph characterizing a
$10 \times 10$ spin glass realization
and of a
bipartite random matrix with $10000$ nodes and $50000$ links.
The quantum ground state energy $E_{0}$ for the spin glass is
$E_{0}/\Gamma=-10.1949$, and for the bipartite graph
$E_{0}/\Gamma=-10.2335$.
The ordinate shows the number of eigenvalues in a
bin of width $0.01$.
The bipartite random matrix has a large gap between the
ground state and first excited state, and, once the gap
is exceeded, a much larger
density of states than the spin glass.
}
\label{fig:random_matrix}
\end{center}
\end{figure}

To obtain context for these results, we interpret the Hamiltonian matrix
for the quantum spin glass as the adjacency matrix of an undirected
bipartite graph~\cite{chartrand85,cvetkovic80,farkas2001} in which each
classical ground state is a node and edges connect every pair of
classical ground states coupled by the quantum term in the Hamiltonian.
The graph is bipartite because the edges connect states that differ by a
single spin reversal, one of which has an even and the other an
odd number of up spins.
The spin glass graphs have a modest number of
disconnected pieces, called clusters~\cite{hartmann9902120}.
Figure~\ref{fig:random_matrix} compares the density of energy levels of
the largest cluster of a $10 \times 10$ spin glass realization (with
17040 nodes and 77684 edges) to the density of energy levels of a
symmetric bipartite random matrix with 10000 nodes and 50000 edges.  The
bipartite random matrix has a large energy gap between the ground state
and first excited state, and once this gap is exceeded the density of
energy levels is much greater than at low energies in the spin glass.
The energy level spacing between the excited states of the random matrix
is approximately inversely proportional to the number of nodes.

\begin{figure}
\begin{center}
\includegraphics[clip,width=0.73\hsize]{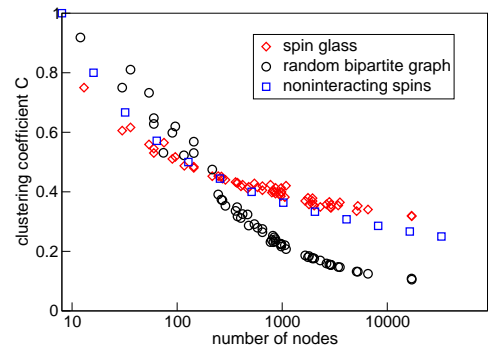}
\caption{
Clustering coefficients $C$ of spin glass graphs,
of bipartite random graphs with the same number of nodes
and edges, and of graphs for noninteracting
quantum spins, versus number of nodes.
The clustering coefficients of the spin glass graphs are significantly
larger than those of random bipartite graphs, and close to those of
graphs for noninteracting spins.
}
\label{fig:clustering}
\end{center} 
\end{figure}

We have compared
other properties~\cite{albert2002} of the graphs underlying
the quantum spin glass
to those of random bipartite graphs.
The degree distribution~\cite{barabasi99} describing the number of links
emanating from
the nodes of the spin glass graphs is even narrower than the Poisson
distribution of a bipartite random graph with the same
mean degree.
Figure~\ref{fig:clustering} shows the
clustering coefficient $C$~\cite{watts98}, which for bipartite
graphs is the probability that two nodes with a
common second neighbor are themselves second neighbors~\cite{newman2001}.
The clustering coefficients of spin glass graphs are
significantly larger than those of bipartite random graphs with the
same number of nodes and edges~\cite{newman2001}, and are
close to those of graphs describing $N$ noninteracting spins, which
have $2^N$ nodes, each node with degree $N$,
and clustering coefficients $C=4/(N+1)$\cite{clustercoefnote}.

Though some statistical properties of
the spin glass graphs are similar to
those of graphs for noninteracting quantum spins, the ground
state of the quantum spin glass differs significantly
from that of
noninteracting quantum spins.
\begin{figure} 
\begin{center}
\includegraphics[clip,width=0.8\hsize]{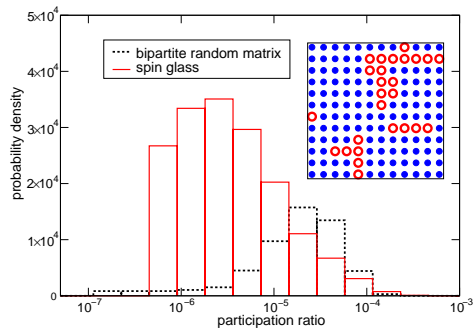}
\caption{
Histogram of the probability distribution
of classical ground
states whose participation ratio $P_\alpha$ is in a given range in
the quantum ground state, on a logarithmic scale.
Here, 
$P_\alpha$ = $|c_{\alpha 0}|^2$, where the $c_{\alpha n}$ are defined
in Eq.~(\protect{\ref{eq:c_definition}}).
The distributions of
participation ratios differ significantly between
the spin glass and the random bipartite matrix,
though in both cases the $P_\alpha$ vary over many orders of
magnitude (note the logarithmic abscissa).
Inset: Spins denoted with open circles have mean magnetization
$\expect{S_i}$ satisfying $|\expect{S_i}| \ne 1$
in the quantum ground state.
}
\label{fig:participation}
\end{center}
\end{figure}
Figure~\ref{fig:participation} (inset) displays the spins
in the quantum ground state of a $12 \times 12$ system that are
fluctuating, in that
$|\expect{S_{iz}}| \ne 1$, where $\expect{S_{iz}}$ is the expectation
value of the $z$th component of the $i^{th}$ spin.
The fluctuating spins form
connected ``bunches''~\cite{Landry2001a} with up to 15 spins,
so a description in terms of noninteracting spins is
not appropriate.
The main panel
of Figure~\ref{fig:participation} shows
the participation ratios $P_\alpha$
of different classical ground states in the quantum ground state,
where $P_\alpha=|c_{\alpha 0}|^2$, with the $c_{\alpha 0}$ defined
in Eq.~(\ref{eq:c_definition}).
The spin glass participation ratios
vary over several orders of magnitude (note the logarithmic
abscissa), while for 
a system of $N$ noninteracting spins, the participation
ratio is $1/2^N$ for all classical configurations. 
Figure~\ref{fig:participation} shows that the participation
ratios of the adjacency matrix of a bipartite random graph also vary
over several orders of magnitude, but with quite different statistics.

We expect the perturbative methods we use to be valid so long
as the energy arising from the quantum perturbation
($\simlt \Gamma N$) is smaller than
the energy gap between the classical ground state and lowest
classical excited states ($=J$), so that
the procedure
is valid only for $\Gamma \simlt J/N$.
Exact diagonalizations of
very small (3x3 and 4x4) systems are consistent with this expectation.

Our results, particularly the low density of well-defined low-frequency
excitations, provide a framework for understanding the generation of
coherent, saturable excitations in a complex quantum spin
liquid~\cite{ghosh2002}.
Saturability arises naturally when the 
density of excitations is low because a frequency that induces a
transition from the ground state to an excited state will not be able
to induce a second transition from the excited state. 
When external excitations couple only two states,
the system will display
coherent oscillations and will have the capability of encoding
phase-coherent information~\cite{ghosh2002}.
An important future goal is to understand
how to implement controlled dynamics involving more
than two energy levels in this strongly interacting system,
as has been achieved in the weakly interacting regime
in the context of
NMR quantum computation~\cite{nmr_qc}.

We have benefited greatly from discussions with J. Brooke,
S. Ghosh, T.F. Rosenbaum, and S.A. Trugman.
This work was supported by the MRSEC program of the National
Science Foundation under Award No. DMR-9808595 at The University
of Chicago and by the National Science Foundation under Award
No. DMR-0209630.  SNC thanks the Aspen Center for Physics
for hospitality during the preparation of this manuscript.

\bibliography{scibib,%
/home/snc/papers/bibfiles/sg,%
/home/snc/papers/bibfiles/spinglass,%
/home/snc/papers/bibfiles/lanczos}

\begin{thebibliography}{32}
\bibitem[{clu()}]{clustercoefnote}
\bibinfo{note}{To calculate the clustering coefficient $C$ for the graph for
  noninteracting quantum spins, consider two distinct second neighbors of a
  given reference state (there are $ \frac{1}{2} \left [ \frac{1}{2} N(N-1)
  \right ] \left [ \frac{1}{2} N(N-1) -1 \right ] $ of such pairs). Note that
  each state in the pair has two spins that differ from the reference state,
  and that they are second neighbors to each other if one of these spins is in
  common. For a given reference state, there are $N(N-1)/2$ ways to choose the
  first state of the pair. If the second state of the pair is a second
  neighbor, then one of the two flipped spins is the same in the second state
  of the pair also. There are $N-2$ possible locations of the second flipped
  spin in the second state, since it can be anywhere except for the locations
  of the two flipped spins of the first state of the pair. Each pair is counted
  twice by this process, so the number of ways to choose a pair of states that
  are second neighbors to each other as well as the reference state is $ \left
  [ \frac{1}{2} \right ] \left [ \frac{1}{2} N(N-1) \right ] \left [ 2 \right ]
  \left [ N-2 \right ], $ and the clustering coefficient $C$ is
  \begin{equation} C = \frac{\left [ \frac{1}{2} \right ] \left [ \frac{1}{2}
  N(N-1) \right ] \left [ 2 \right ] \left [ N-2 \right ]} {\frac{1}{2} \left [
  \frac{1}{2} N(N-1) \right ] \left [ \frac{1}{2} N(N-1) -1 \right ]} =
  \frac{4}{N+1}~. \end{equation}}.
\end{thebibliography}

\end{document}